\documentclass[preprint]{aastex}

\shorttitle {The Mass Function  of M35}
\shortauthors{Barrado y Navascue\'es et al.}

\begin{document}

\title{From the Top to the Bottom of the Main Sequence: A Complete Mass
Function of the Young  Open Cluster M35}

\author{David Barrado y Navascu\'es\altaffilmark{1}} 
\affil{Max-Planck-Institut f\"ur Astronomie, K\"onigstuhl 17,
Heidelberg, D-69117 Germany,
barrado@pollux.ft.uam.es}

\author{John R. Stauffer\altaffilmark{2,3}}
\affil{Harvard--Smithsonian Center for Astrophysics,
       60 Garden St., Cambridge, MA 02138, USA,
jstauffer@cfa.harvard.edu}

\author{Jer\^ome Bouvier\altaffilmark{3}}
\affil{ Laboratoire d'Astrophysique, Observatoire de Grenoble, 
            Universit\'{e} Joseph Fourier, B.P.  53, 38041
Grenoble Cedex 9,
            France, jbouvier@obs.ujf-grenoble.fr}

\and

\author{Eduardo L. Mart\'{\i}n\altaffilmark{3,4}}
\affil{California Institute of Technology, MS 150-21, Pasadena, CA 91125, USA
ege@gps.caltech.edu}

\altaffiltext{1}{Present address: Departamento de F\'{\i}sica Te\'orica, C-XI.
Universidad Aut\'onoma de Madrid, Cantoblanco, E-28049 Madrid, Spain }

\altaffiltext{2}{Visiting Astronomer Kitt Peak National
Observatory. 
KPNO is operated by AURA, Inc., under contract
to the National Science Foundation.}

\altaffiltext{3}{Visiting Astronomer, Canada-France-Hawaii
Telescope }

\altaffiltext{4}{Present address: Institute for Astronomy, 
University of Hawaii, Honolulu, HI96822, USA }

\begin{abstract}

We present very deep and accurate photometry of the  open
cluster M35.
We have observed this association in the Cousins R,I filters,
together with the Johnson  V filter. We have 
covered a region of   27.5$\times$27.5 square arcmin,
 equivalent to a fifth of the total area
of the cluster. The data range from I$_{\rm c}$=12.5 to 23.5
magnitudes, 
and the  color intervals are
0.4$\le$(V--I)$_{\rm c}$$\le$3.0,
0.5$\le$(R--I)$_{\rm c}$$\le$2.5.
Roughly, these values span  from  1.6 M$_\odot$ 
down to the substellar limit, 
in the case of cluster members.
By using the location of the stars on   color-magnitude and
color-color  diagrams, we have selected 
candidate   members of this cluster.
We have merged our sample with previously published 
data and obtained a color-magnitude diagram for the 
complete stellar population of the cluster, covering the 
spectral range early B -- mid M. 
Based on the distribution of field and cluster stars in 
color-magnitude and color-color diagrams,
we estimate that two thirds of these candidates are likely
to be true members of M35.  These stars approximately
double the number of stars identified as candidate  members of this cluster
($\sim$2700).
We provide  the photometry and accurate positions of these stars.
The deep photometry has allowed us to study
the mass segregation within the cluster,   
 the Luminosity Function  and Mass Function.
  We show that in the magnitude range 13 $\le$ I$_c$
$\le$ 22 there is a reduced  mass segregation, in opposition to
what happens to higher mass stars, where the mass segregation is
stronger. The Luminosity Function behaves essentially as the one
characteristic of the Pleiades, presenting a peak at I$_{\rm
c}\sim$19 magnitudes (M$_{\rm I}\sim$9). Combining our  photometry
with previous data corresponding to more massive stars, we find 
 that the Mass Function  increases monotonically, when plotted in
a log-log form, until it reaches $\sim$0.8 M$_\odot$
($\alpha$=2.59).
 It remains shallower for  less massive stars
($\alpha$=0.81 for 0.8--0.2 M$_\odot$), whereas a decrease 
ins observed for stars close to the substellar regime.
These different behaviors suggest that at least three mechanisms
play a role in the formation of stellar and substellar objects. 
The total  mass of the cluster is $\sim$1600 M$_\odot$ in the area
covered by this study. 

\end{abstract}


\section{Introduction}

Open clusters serve as Rosetta stones to study stellar
properties  because they
 provide homogeneous samples of known age, metallicity and
distance.
Therefore, open clusters can be
 used to determine empirically the evolution of different stellar properties
such as 
the activity, rotation, chemical abundances, mass loss, et
cetera.
For low mass stars, however, only
 a handful of open clusters have been studied systematically,
 essentially due to the fact that most  clusters are
far away, with
distance moduli ranging from 6 to 12 magnitudes. With the recent
improvements in
 detectors and with the availability of large ground-based
telescopes, it is
 possible to obtain high quality data for a larger number of
clusters. 

M35 (NGC~2168) is a very interesting open cluster for several
reasons. It is
 moderately nearby, with (m-M)$_0$=9.7 (Vidal 1973). It is very
well
 populated, containing several thousand members
 with a total dynamical  mass estimated between 
1600 and 3200 M$_\odot$ (Leonard \& Merritt 1989).
It is spread over an area of $\sim$1$\times$1 square degrees, making
it a  suitable target for multifiber spectroscopy.
 The contamination by field
 stars is not overwhelming, as happens in some other clusters of
similar  characteristics. 
The reddening is also moderate --E(B--V)=0.17--0.225.
  Perhaps one of the most
interesting properties
is its youth, since it is said to be a cluster coeval with the
Pleiades,
based on  the fact that  their upper-main sequence turn-off locations 
are said to be  at the same position in
 color-magnitude diagrams (Vidal 1973).

Until very recently, only the upper-main sequence  (MS) of M35
had been studied with some detail. Photometric (UBV filters) 
and/or proper motion searches
were carried out by Cuffey (1938), 
Hoag et al. (1961),  Cudworth (1971), Vidal (1973),
McNamara \& Sekiguchi (1986a).
  They reached barely V=15,
about  1.2 M$_\odot$, in the case of M35 members.
 At the present time, different groups have 
conducted several  multi-band  CCD photometry which reach  much
deeper members, such as Sung \& Bessell
(1999=SB1999), von Hippel et al. (2000) and Sarrazine et al. (2000).
The last works present data in the  UBVRI filters, reaching beyond V=19,
whereas Sung \& Bessell (1999) used the  UBVI passbands only, reaching one
magnitude 
fainter.  Sung \& Bessell (1999) were able to estimate the interstellar 
reddening, the distance and the  age
of the cluster, deriving
E(B-V)=0.255$\pm$0.024,
(m-M)$_0$=9.60$\pm$0.10 and 
$\tau$=200$_{-100}^{+200}$ Myr. 
Previously, Sung \& Lee (1992) derived 
a distance modulus (m-M)$_0$=9.3, 
an age $\tau$=85 Myr, and a differential 
internal extinction
E(B-V)=0.26--0.44.
A different estimate of the distance and reddening  
have been published by von Hippel et al. (2000), 
E(B-V)=0.3,
(m-M)$_V$=10.25. These authors also estimated
an age by fitting isochrones, yielding 
$\tau$=100 Myr. However, they inferred that M35 is older than
the Pleiades by comparing in a differential way members of both
clusters. 
Based on an extension of the data presented by von Hippel 
et al. (2000), and assuming a metallicity of
[Fe/H]=--0.21 (Barrado y Navascu\'es et al. 2000a,b),
 Sarrazine et al. (2000)  derived
(m-M)=10.16$\pm$0.10, 
E(B--V)=0.198$\pm$0.008 and 
$\tau$=160$\pm$40 Myr.
This age and the Sung \& Bessell (1999) value are
 significantly  older than the canonical upper-main sequence age
of the Pleiades, 70-100 Myr (Mermilliod 1981;
Meynet, Mermilliod, \& Maeder 1993).
These older ages are in agreement with our own findings
concerning the distribution of rotational velocities and lithium abundances
(Barrado y Navascu\'es et al. 2000a, 2000b).
Based on all the available information, they assumed an
age of $\tau$=175 Myr.
Note that Barrado y Navascu\'es et al. (2000b) have derived
the first metallicity estimate for M35 based on high resolution
spectroscopy, with the resulting value of [Fe/H] = --0.21$\pm$0.10,
 lower than the value 
corresponding to the Pleiades, [Fe/H]=+0.01,
estimated in the same fashion by these studies.
The Pleiades cluster is one of the cornerstones of
 astrophysics, and the comparison of  members of both clusters
can provide
 additional information for understanding their properties. Although
several   stellar characteristics are thought to depend
essentially on age
(rotation, coronal and chromospheric activity, et cetera), this
assumption has
 been recently challenged by Randich \& Schmitt (1995),
comparing the X-ray
 luminosities of Praesepe  and the Hyades, two coeval clusters
(see also the
discussion in Barrado y Navascu\'es et al. 1997, 1998a). Extensive and
detailed 
studies of several clusters of similar age  can help to
understand
this crucial
 problem, as well to clarify the evolution of the stellar
properties when
 comparing clusters of different ages.

We have embarked on an ambitious study  of the  
open cluster M35, combining high resolution spectroscopy to determine
lithium abundances, radial and rotational velocities 
(Barrado y Navascu\'es et al. 2000a, 2000b) with deep imaging
to extend the membership list to nearly the hydrogen burning mass
limit (or beyond) and estimate the cluster Mass Function
(Barrado y Navascu\'es et al. 1998b).
  The current paper
is devoted to the presentation of our deep imaging data,
 to study the complete MF,  the mass segregation within the
cluster and to estimate the total mass of M35.

\section{Observations}

We have carried out photometric observations of the 
young cluster M35  in two different
campaigns. The first run took place in December 11th, 1995, at 
the KPNO\footnote{Kitt Peak National Observatory is operated by
AURA, Inc.,
 Association of Universities to research in Astronomy, under
contract to the
 National  Science Foundation.} 4m telescope. In this case, we
used the prime
focus, a CCD camera, with a  Tek 2048$\times$2048 pixels, yielding  a
field of
 23.2$\times$23.2 square arcmin.  For the second run 
we used the Canada--France-Hawaii Telescope 
 (December 13th, 1996) and the  mosaic
camera  (8000$\times$8000 pixels), which
provided a field of 27.5$\times$27.5 square arcmin.

The observations  were processed using standard
procedures
(bias subtraction, flat-field correction, etc). We used the 
IRAF\footnote{IRAF is distributed 
by National Optical Astronomy Observatories}
 package in the case of the KPNO data. 
Since our aim 
 was to obtain photometry of cool stars over a very large dynamical
range, we collected two sets of images of different 
exposure times during the KPNO run, namely  60 and 600 sec for the
V filter, and  10 and 120 sec for the Cousins I$_{\rm c}$ band
(Cousins 1978). 
 Therefore, we reached 
(V,I$_{\rm c}$)$_{\rm limit}$=(20.2,18.2) for the shallow set and 
(V,I$_{\rm c}$)$_{\rm limit}$=(21.1,19.5) for the deep exposures.
The completeness limits are 
V$_{\rm complete}$=19.0 and
V$_{\rm complete}$=19.75, respectively.
In the case of the CFHT survey,
we collected data in the magnitude interval 
17.5$\le$I$_{\rm c}$$\le$23.5, 
 0.5 $\le$ (R-I)$_{\rm c}$ $\le$ 2.5.
The completeness limit was
I$_{c,{\rm complete}}$=22.1 magnitudes  in this case. 
Errors can be estimated 
as 0.05 magnitudes for each filter at the bright end, degrading to
0.15 magnitudes for the faintest 
objects in each dataset. 

For the KPNO observations, 
the calibration of the data was performed by observing standard stars
in Landolt (1992) --SA 92, 94, 95, 98, 100, 112, and 113-- several times
during the observing run.
We derived our own extinction
coefficients, yielding C$_{\rm V}$=0.192$\pm$0.006 and
C$_{\rm I}$=0.110$\pm$0.003. The calibrated data 
was obtained  by using the expressions:

\begin{equation}
V = v - 3.583 - 0.192 \times airmass - 0.046 \times (v-i) 
\end{equation}

\begin{equation}
I = i - 4.276 - 0.110 \times airmass + 0.012 \times (v-i) 
\end{equation}

\noindent  where V,I denote the calibrated magnitudes, whereas
v,i refer to the instrumental values.

We have compared our final calibrated data with the two previous
studies which are deep enough. The comparison between our data and the
values from Sung \& Bessell (1999) indicates that the I data
are essentially identical, whereas there is a  difference
in the data corresponding to the V, in the sense
V(SB1999)=V(this work)$+$0.03 magnitudes.
Of this difference, 0.01 magnitudes can be attributed
to the difference between the Landolt (1992) photometry
(the values used for the calibration), and the SAAO
photometry, which defines the Cousins system.
Regardless of this   systematic difference,
 our final results
(membership selection and Mass Function analysis)
are not affected by it.

The CFHT observations were carried out
and analyzed  in the same fashion as our observations of the 
Pleiades and Praesepe  (Bouvier et al. 1998).
In all cases the data were recorded under photometric 
conditions and the calibration was performed  
independently from each other using standard 
stars listed in Landolt (1992). Several very red standard
stars were observed to ensure the accurate transformation
from the instrumental system into the Cousins system. 
Visual inspection of the stars in common to the two
set of data (KPNO and CFHT), as
listed in  Table 2,
indicates that the calibrated magnitudes
 are in excellent agreement.

In the case of the KPNO survey, coordinates were derived for all
the detections using the GSC catalog. Positional accuracies are better than
0.5 arcsec.   An empirical confirmation for the accuracy of the
coordinates was provided by the good quality of the spectra we
obtained using our coordinates for a WIYN/HYDRA multi-object
spectroscopy run
(Barrado y Navascu\'es et al. 2000a, 2000b).

\section{The Selection of  Members}

\subsection{A Main Sequence for Cool Stars}

In order to select possible M35 cluster members based on their
location in a color-magnitude diagram (CMD),
 we need to have a reliable main sequence
locus for low mass stars.  Theoretical isochrones (e.g. Baraffe et al. 1998)
do not yet do this optimally for optical colors due to
remaining problems with cool dwarf model  atmospheres 
and inadequate detailed line opacities used to compute
fluxes.  Therefore, we must rely on an empirical main sequence.
We have constructed one using field stars from Leggett (1992) which
fulfill
two conditions: they have accurate parallaxes ($\Pi$), in the sense
$\sigma$($\Pi$)/$\Pi$$\le$0.10,
 and they were not listed as "AB" in
that paper (that is, they are not spectroscopic binaries or
close visual binaries). The selected data come from Luyten
(1972), Giclas et al. (1971), Gliese (1969) and Gliese \&
Jahreiss (1979) --see reference therein for the
original sources of the  photometric data--.
Figure 2 of Bouvier et al. (1998) and Figure 1
of Stauffer et al. (1999) show
this empirical MS when applied to the Pleiades and Alpha Per
open clusters, respectively, in the 
 [I$_{\rm c}$,(R--I)$_{\rm c}$] plane. 
Figure 1 of Barrado y Navascu\'es et al. (1999) displays the
comparison 
between the empirical  MS and field stars from the Leggett (1992)
in the [V,(V-I)$_{\rm c}$] plane. As can be seen, 
the lower envelope is an
appropriate MS. 

For the current paper, we need to extend this MS to bluer
(higher mass) stars. We have done this first by tracing a lower
envelope to M35 members identified by Sung and Bessell (1999) in
the I$_{\rm c}$ magnitude range 8.5 to 14.0.  
Second, we used accurate
radial velocities for 78 of our photometric candidates
down to I$_{\rm c}$=16,    to
determine whether they are indeed members
(Barrado y Navascu\'es et al. 2000b, shown as solid 
circles).
 The loci of those
{\it bona fide} members were used to extend the Sung \&
Bessell
(1999) data and to connect it with the empirical isochrone
 derived from Leggett's data. The final result can be seen in
Figure 1a and Figure 1b.  Note that for the lowest masses,
we would expect M35 members to be above (to the right) of this 
curve because the M35 members will still be contracting to the
main sequence,  assuming the canonical age of the Pleiades. 
However, the low metallicity of M35 --[Fe/H]=--0.21--,
produces a shift of the location of its members compared similar
stars of solar composition and as discussed in Section 1, there
are evidences that M35 is older than the Pleiades.
 At higher masses, our MS curve should serve
as an approximate lower envelope to the distribution of M35
members (e.g. photometric binaries should still scatter up
to 0.75 mag above the curve).
Our empirical MS can be found in Table 1.

\subsection{Color-magnitude and color-color  diagrams}

\subsubsection{KPNO Data}

Figures 1a and 1b depict the photometric VI$_{\rm c}$ data collected
at KPNO. The first panel shows the CMD for the shallow exposure, 
whereas panel  b contains the values for the deep one. In both
cases, limiting  and completeness magnitudes are indicated
as solid and dashed lines. Our empirical MS sequence is also 
shown as a solid line. We used a distance modulus of 
(m-M)$_0$=9.60, a color excess  of 
E(V-I)$_{\rm c}$=0.321 and an interstellar absorption of 
A$_{\rm V}$=0.765 (R$_V$=3.0), which corresponds to the reddening
estimated
by Sung \& Bessell (1999) of E(B-V)=0.255. The probable members
of  M35 (Barrado y Navascu\'es et al. 2000b), observed spectroscopically
($<$RV$>$=--8 km/s), are shown as solid circles.

We have selected as initial possible candidate members of M35 all
those stars which have photometry in a strip described by the
empirical MS and the MS minus 0.75 magnitudes (to allow for binarity)
 and $\pm$0.1  magnitudes, to allow for photometric errors and errors in the
distance  and the reddening.
Hence, the strip is 0.95 magnitudes wide.
Another example of photometric selection of candidate members 
of a cluster can be found in Barrado \& Byrne (1995), in the case NGC5460,
and Barrado y Navascu\'es et al. (2000c) for IC2391.
The selection of M35 candidate members was stopped at
V=21.0 magnitudes. Therefore, we have covered the spectral range
F0--K7 for cluster members.
 Note that despite the lack of separation
between
field stars and the locus of the M35 main sequence, the diagrams
show a relative concentration of objects corresponding to M35,
very clear in Figure 1b in the range 1.5$\le$(V-I)$_{\rm c}$$\le$2.0.

\subsubsection{CFHT Data}

Our initial optical photometric survey for 
members
of M35 was extended by taking longer exposures at the 
Canadian-French-Hawaii telescope. In this case, we obtained 2
exposures in the Cousins R and I filters.
Figure 1c displays the I$_{\rm c}$ versus (R--I)$_{\rm c}$ 
color--magnitude diagram. The objects detected by this survey are
plotted as dots. We have added the empirical MS
for field stars (see  section 3.1),
shifted according to the reddening -E(R--I)$_{\rm c}$=0.178- and the
distance modulus  --(m-M)$_{\rm I}$=10.044--
of M35 (see previous subsection). 
Note that the MS really
traces the relative concentration of very cool stars in M35,
although there is a shift of $\sim$0.1 magnitudes, which can be
attributed to the low metallicity of the cluster. 
Note that the lower, redder end of the CMD corresponds to
objects of M5.5 spectral type. The location of the M35 low mass
candidate members, below our empirical MS,  seems to indicate 
that they are not  pre-MS objects, and therefore, they are
older than $\sim$100 Myr, in agreement with the age
by Sung \& Bessell (1999) and Sarrazine et al. (2000) 
and our previous findings 
(Barrado y Navascu\'es et al. 2000b).

Figure 1c clearly shows two groups of stars: 
 A significant fraction of the stars are on the locus of M35.
 Most of the
detected stars lie between 0.5 $<$ (R--I)$_{\rm c}$ $<$ 1.7,
following a distribution parallel to the 
M35 MS, but 4 magnitudes fainter.
However, there is not a clear gap, empty of stars, 
 between the former (M35 candidate members)
 and the
latter  groups (field stars). A smooth transition is present, indicating that
any list of candidates of M35 presents  contamination by field
stars. In any case, the diagram allows us to perform a
tentative identification of candidates for membership to the
cluster. 
We have identified candidate  members of M35 here
in the same way that we did for the KPNO data, except in this
case  allowing also for the fact that the faintest M35 members
will still be contracting to the MS, and for
uncertainties in our empirical MS, as well as in the cluster
age, distance and reddening.
We have cross-correlated the 
list of stars selected from the VI$_{\rm c}$ plane with the list
coming from the RI$_{\rm c}$ survey, merging both sets of data.
Note that the intersecting magnitude range is only
one magnitude wide, from 17.5 to 18.5  in I$_{\rm c}$.

\subsubsection{The (R-I)$_{\rm c}$--(V-I)$_{\rm c}$ Plane} 

Our two surveys (CFHT and KPNO)
 overlap over a range of one magnitude. Therefore, we have merged
both datasets,   generating a subsample of data with
measurements in three passbands.  Figure 2 shows (R-I)$_{\rm c}$
against (V-I)$_{\rm c}$ for this subgroup.
 We have  included in the figure  the average
behavior  for field stars, after Leggett (1992) and Bessell (1991),
which 
are shown as solid and dashed lines, respectively.
 Very late spectral type members of the young clusters IC2391
(Barrado y Navascu\'es et al. 2000c) are also included as a comparison group.
Based on this diagram,  we have rejected several stars as members
of the cluster, due to their location (cross symbols). The solid
dots represents the  photometric members of the
clusters.

\subsubsection{A List of Members}

Our list of members contains  1945 stars,
with 275 stars previously observed (essentially by McNamara \& 
Sekiguchi 1986a and Sung \& Bessell 1999). Therefore, we present
a list with 1670 new possible photometric members of the cluster, 
doubling the number of candidate  members of the cluster. 
Including all previously known members of M35, there are in total 
 about 2710 identified candidate members.
As a comparison, the Pleiades has 1194 known members
(Pinfield et al. 1998).

Table 2 lists our identifying number for each object and the photometry
in columns \#1 to \#6. Positions are listed in columns 
\#7 and \#8. Finally, cross identification with 
WEBDA database (Mermilliod 1996), McNamara \& Sekiguchi (1986a)
 and Sung \& Bessell
(1999), and membership probability, derived from the proper 
motion, are included from column \#9 to column \#12. Note that 
Mermilliod's identification number in his WEBDA database coincide with 
the identification number of Cuffey (1938) for the first
778 stars. Stars in the  M35 field of view in  WEBDA database
are identified  with numbers from 1 to 4219 (the database includes
1566 entries, some of them are in fact no members).
 We have decided to identify our M35 candidate
members with numbers starting at 5001.

Our final membership assignment  is also listed in Table 2.
Those stars with radial velocity and/or with  proper motions
 in agreement with the cluster average are flagged with ``Y+''
in column \#13. The same criteria were used to reject several
candidates and they have the ``N+'' flag.
Candidate members with positions in the color-color diagram which do not 
correspond  to the cluster locus appear as ``N'', whereas those stars
which only have V,I$_{\rm c}$ and/or R$_{\rm c}$,I$_{\rm c}$ data are listed 
as ``Y''.

Final  color-magnitude diagrams, in absolute I$_{\rm c}$ magnitude
and dereddened colors, are presented in Figure 3a, Figure 3b and Figure 3c.
In all three cases, our empirical main sequence is displayed as a solid line.
In Figure 3a, data from the literature (Sung \& Bessell 1999) appear as 
asterisk symbols (membership probability, from the proper
motions, larger than 0.80, ``Y$+$''). Our {\it bona fide} members, selected
from their radial velocity (Barrado y Navascu\'es et al. 2000b),
 are shown as solid circles (``Y$+$''). Other
photometric members are displayed as open circles (``Y''), whereas
non-members appear as crosses (``N'', ``N$+$'').
Figure 3b displays the photometric candidate members from the CFHT survey
--I$_{\rm c}$ versus (R--I)$_{\rm c}$.
A complete color-magnitude diagram for all {\it bona fide}
 and candidate members of M35 is shown in Figure 3c, stretching 
from M(I$_{\rm c}$)=--1.5 up to M(I$_{\rm c}$)=13.0, and covering
from early B down to M5.5 spectral types.
It includes our own candidate members and those candidate members
from Sung \& Bessell (1999) which are probable members,  based on
their proper motions, as derived from McNamara \& Sekiguchi (1986a).
Note that in order to show our CFHT data at the red, faint end, 
we derived  (V--I)$_{\rm c}$ colors from  (R--I)$_{\rm c}$
and  our empirical MS.
The star HD41996, which is a likely member based on its proper motion
and  has already evolved off the MS, is out of the limits of the
figure.

\subsubsection{Brown Dwarfs in the cluster}

Our fainter candidates reach I$_{\rm c}$$\sim$23.0, which 
corresponds to M(I$_{\rm c}$)$\sim$13.0 magnitudes. Following
Stauffer, Schultz \& Kirkpatrick (1998) and the theoretical
models of Baraffe  (1998, priv. communication), the substellar
limit should be, for an age of 125--200 Myr, at
I$_{\rm c}$(BD)$\simeq$22.1--22.6, and the lithium
depletion boundary (see Stauffer \& Barrado y Navascu\'es  2000) at
I$_{\rm c}$(LDB)$\simeq$22.3--23.2 magnitudes.
 Therefore, we have 
reached the substellar domain in M35, and found 
a sample of the farthest brown dwarf (BD) candidates discovered
so far. In total, we have 65 objects with masses in the
range 0.075--0.055 M$_\odot$, based on a 100 Myr isochrone
from D'Antona \& Mazzitelli (1994).
 Note that if the masses of cluster brown  dwarfs 
sum up about  a few percent of the total mass of the association, 
it would mean that M35 has a minimum of 600--1000 substellar objects.
For instance,  brown dwarf searches in 
the Pleiades  have estimate that the total contribution 
of BD to the mass of the cluster can be 5\% (Bouvier et al. 1998) 
or 3\% (Hodgkin \& Jameson 2000).

\subsection{Contamination by field stars}

The relatively smooth transition between the concentration of
field stars, fainter and/or bluer than M35, and this cluster,
indicates that our list of member candidates  is
contaminated by field stars.
However, note that in the case of the V,(V-I)$_{\rm c}$ diagram, the
gap between the M35 main sequence
and the bulk of the field stars increase with the magnitude:
 the separation in color
is smaller at the upper part of the diagram and, therefore,  the
number of spurious members should be larger for stars of spectral
type G and early K.  
Moreover, the separation in color between the field stars and 
the cluster locus is larger in the case of the I,(R--I)$_{\rm c}$
diagram, a fact which indicates that 
it is easier to estimate  the contamination by field
stars  in this case than  from the 
V,(V-I)$_{\rm c}$ diagram.

In the case of the brighter stars presented in this work, 
which come from the V,(V-I)$_{\rm c}$ CMD, 
we have estimated the contamination by field stars using
two different methods: 
(i) Some of the stars have been discovered previously, and proper
motions and their associated membership probability are listed
by the authors (Cudworth 1971; McNamara \& Sekiguchi 1986a). 
They have magnitudes in the range 13.7$\le$V$\le$14.9, which
correspond to F2-G2 spectral types. We have
compared the number of stars present in our survey having 
probability less  than 50\% with the total number,
 and assumed that  this ratio is  the contamination rate. The result
is 37\%. If a more demanding criterion is selected (such as
membership probability less than 80\%), the pollution 
rate does not change much (53\%).
(ii) We have previously derived radial velocities for a subsample
of these stars (Barrado y Navascu\'es et al. 2000b).
They have magnitudes in the range 14.5$\le$V$\le$17.5, 
corresponding to F9-K5 spectral types.
 We have used the radial velocity as a membership
criterion. The  pollution rate derived from the comparison between
the {\it bona fide} members and non-members
is 30\%. 
These numbers confirm our initial conclusion that contamination is stronger
in the upper Main Sequence (for F and G stars) than for K members.

In the case of the deep survey from CFHT,
we  have divided the I$_c$,(R--I)$_c$ color-magnitude diagram in
boxes of size
0.5 magnitude by 0.1  magnitude in  I$_{\rm c}$ and (R--I)$_{\rm c}$,
respectively. We have constructed diagrams representing the
number of stars in each {\it i} box --N$_{i,*}$-- against (R-I)$_{\rm c}$ for
a given
I$_{\rm c}$ interval. Figure 4 shows one  such diagram,
corresponding to the interval 18.0 $\le$ I$_{\rm c}$ $<$ 18.5
magnitudes. 
The numbers of stars are shown as solid circles.
Two peaks are apparent, one near (R--I)$_{\rm c}$=0.65
and the other close to (R--I)$_{\rm c}$=1.35, 
the last one  corresponding to the location of
M35. Then, we have counted the number of stars in the box
corresponding to the M35 peak, the box immediately bluer, and
the two redder boxes,  delineated  in Figure 4 by the horizontal line
segment. The total number of stars in all four boxes,
 N$_{i,*}$(I$_{\rm c}$),  is the
initial number of members of M35 in the considered interval of
I$_{\rm c}$.
In order to remove the contamination due to field stars, we have
fitted three Gaussian curves  to N$_{i,*}$(I$_{\rm c}$), one
corresponding to the bulk of the field stars, the other to M35,
and the third to the transition between the former two. There is
no physical reason for this procedure, but the shape of any
particular peak is pseudo-Gaussian, and we know that 
 N$_{i,*}$ --the number of stars in each box--
 has to be null on both sides of Figure 4 or the equivalent in
other
I$_{\rm c}$ ranges. The computed distribution of field stars is
 shown as a dashed line in the figure. Note the excellent agreement 
between the data and the fit for the field stars.
We have subtracted the computed number of field stars for each
box from  the measured number, for the boxes corresponding to the 
location of M35. This value, N$_{i,*}^{\rm M35}$(I$_{\rm c}$),
 is the number of  cluster members  and 
is listed in  column \#5 of Table 3 (in the case of the CFHT data).

Table 3 contains the ratio between the spurious  members of the cluster
and the total total number of stars --members$+$field stars--,
per  bin 0.5 magnitude  wide in I$_{\rm c}$ (column \#4).
 This is a preliminary
result, and only more accurate methods, such as proper motion
measurements or accurate radial velocities,
 can guarantee the membership of a particular star
or group of stars.

\section{Mass segregation in the center of the cluster}

Stellar clusters are not stable from the gravitational point of
view. They disipate as they orbit around the Galaxy,
due to the gravitational interaction with the spiral arms, interstellar 
clouds, other clusters, et cetera. In
addition, since they behave  as a N-body problem, they try to find
gravitational equilibrium by
 expelling members, which tend to be the less massive, while
the most massive stars move toward  the center. Therefore,
 mass segregation appears in relaxed clusters.

M35 has an angular diameter of, at least, 33 arcmin, since proper
motion
members have been detected at that distance (Cudworth 1971).
 In fact, Leonard \&
and Merritt (1989) stated that the tidal radius could reach between
33 and 66 arcmin, extending the cluster even further.  Our survey
covers an area of 27.6$\times$27.6 arcmin, a significant fraction
of the total area in the core.
 In this central part of the cluster, we have
 found strong evidence of a limited mass segregation, in a
specific way. This mass segregation appears
when we compare our sample with the sample of more massive
stars   from McNamara \& Sekiguchi (1986a).
McNamara \& Sekiguchi (1986b) found that essentially
almost 100\% of the proper motion members are within an angular
radius of 
20 arcmin, regardless of the mass (6.0$<$Mass$<$1.2 M$_\odot$), based
on data from
 Cudworth (1971). They also established
 that there is mass segregation in this mass range, using either
Cudworth's  or their own data (see their Figure 1).
Similar conclusion has been reached by von Hippel et al. (2000),
for stars brighter than V=19 magnitudes.
However, our data, composed of less massive stars, are distributed
homogeneously on the  surveyed sky section,
without any clear decrease in the number of stars toward the
borders, 
indicating that the distribution of low mass stars could be extended
far beyond the 20 arcmin radius.

To illustrate this situation further, 
Figure 5 displays the cumulative distribution of stars versus
the angular distance (in arcmin, bottom of the diagram) or radius
(in parsecs, top of the diagram).
Open circles represent the stars from Sung \& Bessell (1999)
which have I$_c$$\le$13 magnitudes, whereas  solid and open
triangles represent data from our survey 
(13$<$I$_c$$\le$17.5 and  17.5$<$I$_c$$\le$21.5, respectively).
The mass segregation for the brighter members seems obvious.
However, the two fainter subsamples resemble each other, except
in the interval 3.5--7 arcmin, where a small difference appears.
Moreover, the distributions in the intervals I$_c$=17.5--19.5 and
I$_c$=19.5--21.5 are almost identical.
However, some mass segregation clould be present at the very bottom
of the MS, around the substellar limit
(see Subsection 5.2.4).
These facts indicate that the cluster is not fully relaxed.

\section{Luminosity and Mass Functions}

\subsection{The Luminosity Function}

Figure 6 shows the Luminosity Function of M35, once the spurious
members were removed.
Solid circles represent data coming from the KPNO survey, whereas
open circles correspond to the fainter data, collected at CFHT.
 We have included the errors as vertical bars (see section 5.1.1).
 Note that apparent magnitudes are
indicated with the bottom of the x-axis, whereas absolute magnitudes
appear at  the top. 
 The distribution   peaks around I$_{\rm c}$=19, as happens in the
Pleiades (Zapatero-Osorio 1997). Fainter than this magnitude,
 the number of stars decreases with magnitude and a strong
contamination by field stars appears. Our sample is complete
up to I$_{\rm c}$=22.0 magnitudes.  Therefore, the last point
is uncertain and represents only an upper limit.
We have not considered the effect of binarity when computing the LF
(or Mass Function).

There is an additional remark regarding  the position of M35 in the CMD.
We located the ZAMS in the diagram using the published values
in the literature and appropriate conversions to the photometric
 bands we have used. However, 
we did not use this information when counting the number
of stars in M35. Therefore, our method is totally independent of distance
modulus
and reddening. However, we have obtained that the maximum density
of stars,
given a (R--I)$_{\rm c}$ value, appears $\sim$0.1 magnitudes  below 
the ZAMS, as a visual inspection indicates in Figure 3b.
This seems to be an effect of the low metallicity of M35.

\subsubsection{Estimation of the errors}

To estimate the uncertainties in the Luminosity Function, we took
into account the error in the photometry and those produced by 
the way we removed the contamination by field stars. The error in
I$_{\rm c}$ is 0.05 mag. Since the bin size is 0.5 magnitudes,
20\% of the number of stars
 --N$_{i,*}$(I$_{\rm c}$)-- in a particular {\it i} box could be
assigned to a nearby box ({\it i+1} or {\it i-1}).
 In addition, we only  took into
account the four boxes around the location of M35
(two redder and one bluer, in addition to the box located
at the position of the cluster MS). The error in the
removal of the field stars, $\Delta$N$_{\rm field}$, can be
estimated as the square root of the sum of the the number of stars
in
the external boxes delimiting the cluster, N$_{i,*}^{\rm inf}$ and
N$_{i,*}^{\rm sup}$ in the case of the faint stars or as 30-40\%
for the brighter stars of our sample. Therefore, the total error
is:

\begin{equation}
\Delta N_{i,*}^{\rm M35}(I_{\rm c}) = 
 \{0.2^2\times[(N_{i+1,*}(I_{\rm c})-N_{i,*}(I_{\rm c}))^2
+(N_{i,*}(I_{\rm c})-N_{i-1,*}(I_{\rm c}))^2] + 
(\Delta N_{\rm field})^2   \}^{0.5}
\end{equation} 

The values can be found in column \#5 of Table 3.

\subsection{The Mass Function}

\subsubsection{Estimating the stellar mass}

We estimated the masses using the I$_{\rm c}$ magnitudes.
We used  several isochrones from different
groups, namely D'Antona \& Mazzitelli (1994), 
D'Antona \& Mazzitelli (1997), 
Baraffe et al. (1998) and
 Siess et al. (2000). In the first case, 
luminosities were converted into bolometric magnitudes and these into
I$_c$ magnitudes via bolometric corrections of  Monet et al.
(1992) and the distance modulus of the cluster.
Baraffe et al. (1998) and Siess et al. (2000) provide
colors and magnitudes.
Note that the mass range of the evolutionary tracks is different
for each group and that in the last cases they do not cover 
the whole range of this study.

Figure 7a displays a comparison between the Mass Functions 
computed with 100 Myr isochrones from each model.
From top to bottom, D'Antona \& Mazzitelli (1994)
--with the magnitude-mass relation quoted above--,
D'Antona \& Mazzitelli (1994)
--where we estimated I magnitudes from luminosities
based on Siess et al. (2000) data--,
D'Antona \& Mazzitelli (1997),
Baraffe et al. (1998) and
Siess et al. (2000).
For each MF, we have fitted a 
Salpeter  law $\psi$(M)=k$\times$M$^{-\alpha}$, in the mass range
1--0.2 M$_\odot$. The fits are shown as solid lines.
The power law indices are, respectively,
$\alpha$= 0.98, 1.10, 1.13, 0.94 and 1.14.
  For clarity, we have added 
shifts of 0.4 dex to the MFs compared to the previous one
above it.
As the figure clearly shows, the general 
trend is very similar in each case, independent
of the model. Below 0.2 M$_\odot$  there is a substantial
decrease in the number of objects. Note that our completeness
limit is below this value, about 0.1--0.08  M$_\odot$, depending
 on the model. Therefore, 
this structure in the MF indicates that the efficiency
of the fragmentation process of the original cloud
of the cluster was low below  0.2 M$_\odot$, producing a 
smaller number of objects compare with stars more massive
than 0.2 M$_\odot$. Another possibility is that a fraction 
of original low mass objects have been expelled during the
cluster lifetime (i.e., mass segregation has occurred).

\subsubsection{The effect of the age on the Mass Function.}

As stated before, there is a considerable disagreement 
on the age of M35. Initial estimates were close to 70--100
Myr, the age of the Pleiades (Vidal 1973) and they were
based  on photographic and photoelectric photometry.
More recent CCD photometry (Sung \& Bessell 1999, von Hippel et al. 2000)
indicates that the cluster is somewhat older than the Pleiades,
up to 200 Myr.
Our own estimate, based on rotational velocity and lithium 
abundance of {\it bona fide} members  of the cluster, 
is 175 Myr. We have tried to avoid this controversial issue
by deriving  the Mass Function of the cluster with isochrones
of different ages. Figure 7b shows the results.
Isochrones corresponding to ages of 100, 125, 160 and 200 Myr
and models from Baraffe et al. (1998)
-- dashed lines--  and  
Siess et al. (2000)
-- solid lines-- were used to derive the MFs.
Arbitrary shifts of 0.3 and 0.5 dex were used to display different MFs.
Again, the results are very similar in all cases. The position
of the maximum depends slightly on the age of the isochrone and can 
be moved from 0.2 M$_\odot$ --100 Myr-- to 0.25  M$_\odot$
--200 Myr--. In general, the effect is a shift of the Mass Function
to the left side of the diagram, but keeping the shape of the MF 
unchanged. Therefore, 
within the uncertainty in age, 
we can ignore the effect of age on the MF.

\subsubsection{The complete Mass Function for M35}

A complete Mass Function of the open cluster M35 is 
displayed in  Figure 8.
In panels a and b, 
 M35 data are shown as crosses (photometry from Sung
\& Bessell 1999, corrected by mass segregation), solid circles
(VI$_c$ data from  KPNO) and open circles (RI$_c$ data from CFHT). 
Panel a also includes 
  a Salpeter  law $\psi$(M)=k$\times$M$^{-\alpha}$
 ($\alpha$=2.35, note that only stars more massive than
0.6 M$_\odot$ were used for the fit),
 whereas panels b and c show the M35 Mass
Function in a log-log version.
For comparison purposes, we display on this last two figures data from
the Pleiades open cluster, as a dashed
lines (Figure 8b) or thin solid lines (Figure 8c).
 Error bars are included for
our M35 data. 

Figure 8a clearly shows that a Salpeter  law does not 
fit the whole Mass Function of the cluster. 
We have fitted three different power laws, yielding
$\alpha$=$+$2.59$\pm$0.04 for the mass range 6--0.8 M$_\odot$,
$\alpha$=$+$0.81$\pm$0.02 for 0.8--0.2 M$_\odot$,
$\alpha$=--0.88$\pm$0.12 for 0.2--0.08 M$_\odot$ (Figure 8b).
Note that the mass limits are different to those 
used in previous subsections and in Figure 7a.
For the upper part of the main sequence of M35, 
 Leonard \& Merritt (1989)
obtained $\alpha$=$+$2.7$\pm$0.4 (6--1  M$_\odot$).
In the case of M35 stars slightly cooler, Sung \& Bessell (1999)
derived $\alpha$=$+$2.1$\pm$0.3 (4.4--0.7 M$_\odot$).
Our M35 Mass Function 
 can be compared with the MF for the Pleiades, computed
for mass below $\sim$0.5 M$_\odot$,
by Mart\'{\i}n et al. (1998)
 --$\alpha$=$+$1.00$\pm$0.15, 0.4-0.04 M$_\odot$--,
Bouvier et al. (1998) --$\alpha$=$+$0.6, 0.4-0.06 M$_\odot$--,
Hambly et al. (1999) --$\alpha$=$+$0.7, 0.6-0.06 M$_\odot$--,
Hodgkin \& Jameson (2000) --$\alpha$=$+$0.8, 0.50--0.04 M$_\odot$--,
and Mart\'{\i}n et al. (2000) --$\alpha$=$+$0.53, 0.4-0.06 M$_\odot$--.
All these MFs, arbitrarily scaled, 
 are shown in Figure 8b as solid and dashed lines.
In all cases, the power law indices are quite similar, 
when we compare similar mass ranges.
Note that our statistics are much better due to the richness of the
M35 cluster compared with the Pleiades.

A comparison between a complete Mass Functions of M35 (this work)
and the Pleiades (Hambly et al. 1999) can be found  in Figure 8c.
Solid circles correspond to M35 data, whereas solid triangles 
represent the Pleiades. Note the remarkable similarity
between both clusters, despite the different age, metallicity and 
total mass (see next subsection).
We interpret the difference in the mass range 0.3--0.08 M$_\sun$ 
essentially as arising from the different binning between  our data and 
the dataset by Hambly et al. (1999). Since we have 
a large number of candidate members at the bottom of the main sequence,
we have been able to discriminate smaller bins close to the
substellar limit, i.e., we have four points
in the area around 0.1 M$_\odot$, where the Pleiades data presents
an apparent  gap, due to its wide bins.
 Therefore, we are able to distinguish  more structure, and the 
real decrease of the MF in this mass range.
However, we cannot rule out the possibility of a real difference in the MFs,
even a mass segregation in the case of M35 for very low
mass objects. For instance, BD searches in the Hyades (600-800 Myr) 
have not produced any positive identification
(Gizis et al. 1999).
Figure 8c also includes data from Gould et al. (1997), corresponding
to  field stars. This dataset presents a flat MF in the
mass range 0.5--0.08 M$_\odot$, with a step and narrow 
decrease at 0.1 M$_\odot$.
As shown by other authors in the Pleiades and in the
very young cluster (1-5Myr) associated to  sigma
Orionis  (B\'ejar  et al. 2000), the MF  increases again
below the substellar limit.

In summary, it seems that the Mass Function presents four 
different types of behavior:
From the top of the main sequence down to $\sim$0.8 M$_\odot$,
the MF is very step, with a power index $\alpha$$\sim$$+$2.6
(where $\psi$(M)=k$\times$M$^{-\alpha}$).
In the mass range 0.8--0.2 M$_\odot$, the MF is shallower, with a index 
close to   $\alpha$$\sim$$+$0.9. Between this point  and the substellar limit, 
a real decreases in the M35
Mass Function takes place. The index can be
 evaluated as $\alpha$=--0.8. Finally, as shown by B\'ejar et al.
(2000) in the whole substellar domain,
 the tendency is inverted and a new increment in the Mass  
Function appears, with $\alpha$=$+$0.8.
These different  power law indices suggest that several
mechanism (at least three) are acting during the fragmentation and 
collapse of the original cloud, competing with each other.

\subsubsection{The total mass of the cluster}

The total dynamical mass of M35 has been estimated
as 1600-3200 M$_\odot$
 (Leonard \& Merritt 1989). Using mass function arguments, Sung \& Bessell
(1999) computed  a total mass of 1660 M$_\odot$.
 The average mass of a star
is 0.4 M$_\odot$, which means M35 should contain from 4000 to 8000
members. 
Prior to our study, about 1500 candidate  members of 
M35 were known. We have proposed another $\sim$1700, which have a
contamination rate close to 30\%. In total, about 2700 members
of M35 have been identified so far. This fact seems to indicate that the 
total mass of the cluster is closer to the 1600 M$_\odot$ 
estimate than to the 3200 M$_\odot$.
In fact, our own estimates are in excellent agreement with this
rough estimate.

 We have used the complete dataset displayed in Figure
3c to derive the total mass of the cluster. This sample includes
our new candidate members as well as all photometric members discovered 
previously. Masses were derived using a 100 Myr isochrone (D'Antona \&
Mazzitelli 1994).

 Figure 9a, 9b and 9c represent the cumulative 
total mass distribution against the  M(I$_{\rm c}$) absolute  magnitude, 
(V--I)$_{\rm c,0}$ unreddened color index and
the individual mass, respectively. 
In the case of Figures 9a and c, we derived masses from the 
I magnitudes, whereas in the case of Figure 9b, we used the 
color. Essentially, results are very similar.
The vertical long-dashed straight segment represents the location of the
substellar limit (Figure 9b and c).
The other three curves  represent different calculations of the
cumulative total mass of the cluster:
The dotted line correspond to all probable members and candidate members.
Therefore, it is affected by the field star pollution.  
The thick solid line represents the optimal estimate, where
we included  all  probable
members (using proper motions and radial velocities)
and candidate members (our candidate members and previously known
candidate members without known proper motions),
but removing the spurious members ($\sim$30\%).
Finally, the short-dashed   solid line corresponds
to the probable members and candidate members once the field contamination
has been removed. In this last case, we have not included candidate 
members from previous surveys (i.e., members with no additional information
regarding the proper motion and/or the radial velocity) and the final value
represents the minimum total mass of the cluster. 
The location of this value  is indicate with a horizontal segment
in Figure 9c.

Figure 9c clearly shows that the main contribution to the total
mass of the cluster is produced by stars in the mass range 3--0.5
M$_\odot$, with a inflection point at 0.8 M$_\odot$.
The 0.8 M$_\odot$ point is also the location of another inflection point in 
the cluster  MF, where the spectral index of the power law changes from
$\alpha$=$+$2.59 to $\alpha$=$+$0.81.

The total mass within the area covered by this survey is
$\sim$2100, 1600 and 1100 M$_\odot$, for each case described above,
where the value in the middle (1600 M$_\odot$) can be considered
as a good estimate of the total mass of the open cluster M35 in the 
considered region. 
As a comparison, the {\it total}
 mass of the Pleiades has been estimated as 1000--1200 M$_\odot$
(Meusinger et al. 1996) or 735 M$_\odot$, (Pinfield et al. 1998),
and the total mass of Praesepe is about 626 M$_\odot$
(Holland et al. 2000).
Note that some 
 remaining M35 members should be located in the corona of the cluster, where
deep photometric and proper motion surveys have  not been carried out.
Therefore, the total mass of M35 should be considerably larger than the mass
 of the Pleiades.
Pinfield et al. (1998) have estimated that the average mass 
of a Pleiades star is $<$mass$>$=0.616 M$_\odot$. In the case of
M35, the average mass is $<$mass$>$=0.602 M$_\odot$, very similar
despite the difference in the number of stars in each cluster
(725 against 2713, respectively).

\section{Summary and Conclusions}

We have obtained deep V(RI)$_{\rm c}$ photometry of the open
cluster M35, reaching
much fainter stars than the previous surveys. The comparison of
the location
of the stars in color--magnitude and color-color diagrams allowed
us to establish the preliminary
membership of the detected stars. We have also estimated the  
 contamination by field stars.
We have analyzed this accurate, deep  photometry of M35.
 Our survey has shown that in the studied color range there is a weak
mass  segregation, in opposition to the stronger segregation
present for more massive stars. These data have allowed us to
establish the Luminosity Funtion and 
Mass Function of the low and very low mass components of the
clusters.
The LF presents a peak at I$_{\rm c}$=19 magnitudes, corresponding
to  a mass of $\sim$0.3 M$_\odot$. This result is very similar to the
Pleiades' LF.
By merging our own dataset with data from  previous studies of the cluster,
which correspond to more massive stars, we have been able to obtain a 
complete mass function for the stellar component of the cluster,
covering
the mass range 6.0 $\le$ Mass $\le$ 0.08 M$_\odot$.
When expressed in a  log-log form, the MF increases monotonically
in the
6-0.8 M$_\odot$ domain with a power index of $\alpha$=$+$2.59.
After this point, the power index is reduced significantly, to
a value of $\alpha$=$+$0.81. 
Below this point, the behavior changes and the power
law index become negative ($\alpha$=--0.8). 
Finally, we have estimated the total mass of the cluster
in the area were our data were collected. After taking into account 
the pollution by field stars, the total mass
contained in the central part of the cluster is $\sim$1600 M$_\odot$.
It would be very interesting to collect deeper photometry
of the cluster, reaching well below the
substellar limit, to verify whether the M35 MF behaves as the Pleiades 
and sigma Orionis Cluster
or if a significant fraction of the M35 BDs have already been dissipated
from the cluster.


\acknowledgements

DBN thanks the  {\it ``Instituto Astrof\'{\i}sico de Canarias''}
and {\it ``Ministerio de Educaci\'on y Cultura''}
 (Spain), and the {\it ``Deutsche Forschungsgemeinschaft''}
(Germany) for   their fellowship. JRS acknowledges
support from NASA Grant NAGW-2698. This work has been partially suported by 
Spanish {\it ``Plan Nacional del Espacio''}, under grant ESP98--1339-CO2.
 The comparison with previously published
data has been much easier due to the work done by J.-C.
Mermilliod,
via his WEBDA database. We greatly appreciate this contribution
to the astronomical  community.
We thank the referee, M.S. Bessell, for his multiple and
useful comments.

\newpage

\figcaption{
{\bf a}  KPNO bright data: V magnitude against (V-I)$_{\rm c}$ color, 
shallow exposure.
Solid circles indicate the position of {\it bona fide} M35 members, based
on radial velocity data (Barrado y Navascu\'es et al. 2000a,b). The dashed and
solid lines represent the completeness and detection limits, respectively.
{\bf b} KPNO bright data: V magnitude against (V-I)$_{\rm c}$ color,
 deep exposure. Symbols as Figure 1a.
{\bf c} CFHT deep data:
 I$_{\rm c}$ magnitude against (R-I)$_{\rm c}$ color
index. The dashed line represent the completeness limit.
\label{fig1}}

\figcaption{Color-color magnitude diagram for M35. Solid circles
represent the the M35 candidates selected as photometric
members, whereas the crosses indicate the position of the rejected
 candidates  (see text).
As a comparison, we show  IC2391 probable members  (asterisks) 
from Barrado y Navascu\'es et al. (2000c).
The solid line represents the relation derived using Leggett
(1992) data, whereas the dashed line corresponds to the
 relation for M dwarfs published by Bessell (1991).
\label{fig2}}

\figcaption{
{\bf a} Final color-magnitude diagram for M35, KPNO data. Radial
velocity members appear as solid circles (from Barrado y Navascu\'es
et al. 2000a,b), whereas other 
candidate members are displayed as open circles. Rejected candidate
members, based on Figure 2, appear as crosses. Members from
Sung \& Bessell (1999) are shown as asterisk symbols.
{\bf b} Final M35 color-magnitude diagram, CFHT data.
Solid circle represent stars with V(RI)$_{\rm c}$ data  in agreement
with membership, whereas open circles represent candidate members
which only have RI photometry.
{\bf c} A Complete color-magnitude diagram for the open cluster  M35.
The diagram includes all our 
candidate members and the {\it bona fide} members 
by Sung \& Bessell  (1999).
Note that for the faint end (V-I)$\geq$2.0, most of the color indices 
were derived from the (R--I), using  our empirical MS.
\label{fig3}}

\figcaption{Estimation of the pollution by field stars
fitting Gaussian curves. We show one of the cuts
along the I$_c$ axis in the color-magnitude diagram.
Three Gaussian curves were fitted as described in the 
text. The dotted line represents the fitted
field star contribution.
\label{fig4}}

\figcaption{Mass segregation in the cluster. Open circles, 
solid triangles and open triangles represent data from Sung \&
Bessell
(1999), KPNO and CFHT, respectively.
\label{fig5}}

\figcaption{Luminosity Function. Solid and open circles represent
data from KPNO data and CFHT data, respectively.
\label{fig6}}

\figcaption{
{\bf a} Comparison of several M35 MFs derived with 
different models.  The power law index
for each Mass Function is  shown.
In all cases, 100 Myr isochrones were used.
For clarity, we have added arbitrary shifts of 0.4 dex, respect the curve
inmediately above. 
{\bf b} Comparison of several M35 MFs derived with 
different isochrones and models.
Models  from Siess et al. (2000) appear as solid lines and circles,
whereas isochrones from Baraffe et al. (1998) are
plotted as dashed lines and open circles. 
For clarity, we have added arbitrary shifts of 0.3 and 0.5 dex. 
\label{fig7}}

\figcaption{
{\bf a} Complete Mass Function of M35. Crosses, solid and 
open circles represent
data from Leonard \& Merritt (1989), KPNO data and CFHT data, respectively. 
The solid line is a MF with a Salpeter index ($\alpha$=2.35),
for stars more massive than 0.6 M$_\odot$.
{\bf b} Complete Mass Function of M35 in logarithmic form
(thick solid line). The results from Leonard \& Merritt (1989)
and Sung \& Bessell (1999) are displayed as thin solid lines.
Different Pleiades MFs, scaled with arbitrary shifs, 
 are shown as dashed lines. From top to bottom:
Mart\'{\i}n et al. (1998),
Hodgkin \& Jameson (2000), 
Hambly et al. (1999),
Bouvier et al. (1998) and Mart\'{\i}n et al. (2000).
The spectral power indices of the MF are indicated
inside the parenthesis. 
{\bf c} Complete Mass Function of M35 in logarithmic form
(solid circles). The best fit is shown as a thick solid line.
 As comparison, the complete Mass Funtion
of the Pleiades is shown as solid triangles (Hambly et al. 1999),
with an arbitrary shift of 0.4 dex.
The fits and the power law indices within their range 
are also plotted in the Pleiades case.
Data from Gould et al. (1997), corresponding to field
stars, are included as crosses.
\label{fig8}}

\figcaption{{\bf a} Cumulative total mass of the cluster versus
the apparent  I$_{\rm c}$
 magnitude (top) or the absolute M(I$_{\rm c}$) magnitude (bottom).
Different lines represent several estimates of the total mass
of the cluster (see text).
Stellar masses were derived from the I$_{\rm c}$ magnitudes.
{\bf b} Cumulative total mass of the cluster versus
the dereddened (V--I)$_{\rm c}$ color index.
Different lines represent several estimates of the total mass
of the cluster (see text).
 The vertical long-dashed line
represent the location of the  substellar border at 
$\sim$0.075   M$_\odot$ for a 100 Myr old object.
Stellar masses were derived from the (V--I)$_{\rm c}$ color indices.
{\bf c} Cumulative total mass distribution of the cluster versus
the individual mass. The minimum total mass for this last 
case is indicated with a horizontal segment. Symbols as in Figure 8b.
Stellar masses were derived from the I$_{\rm c}$ magnitudes.
\label{fig9}}

\newpage

\setcounter{table}{0}
\begin{table}
\caption[]{Empirical M35 main sequence. We used 
(m-M)$_0$=9.60, E(B--V)=0.255,
E(V--I)$_{\rm c}$=0.321 and 
E(R--I)$_{\rm c}$=0.178, with R$_{\rm V}$=3.0. }
\begin{tabular}{ccc}
\hline
 V    & (V--I)$_c$& (R--I)$_c$  \\
 (1)  &   (2)     &   (3)       \\
\hline
12.544 & 0.410 &  --      \\          
12.919 & 0.519 &  --      \\         
13.391 & 0.640 &  0.317   \\          
13.513 & 0.675 &  0.336   \\          
13.864 & 0.761 &  0.363   \\          
14.083 & 0.801 &  0.378   \\          
14.368 & 0.827 &  0.389   \\          
14.621 & 0.855 &  0.401   \\          
14.922 & 0.901 &  0.419   \\          
15.468 & 0.961 &  0.440   \\          
15.969 & 1.066 &  0.479   \\          
16.228 & 1.115 &  0.505   \\          
16.828 & 1.245 &  0.591   \\          
17.786 & 1.490 &  0.697   \\            
18.676 & 1.806 &  0.916   \\          
19.952 & 2.294 &  1.278   \\            
20.741 & 2.540 &  1.396   \\          
22.419 & 2.986 &  1.695   \\          
23.244 & 3.111 &  1.772   \\          
24.815 & 3.499 &  1.983   \\          
25.458 & 3.654 &  2.068   \\     
26.100 & 3.946 &  2.208   \\     
26.984 & 4.270 &  2.348   \\     
27.477 & 4.603 &  2.478   \\     
28.194 & 4.940 &  2.598   \\   
\hline
\end{tabular}
$\,$\\
\end{table}

\setcounter{table}{2}
\begin{table*}
\caption[]{ Data for the LM and MF for the low mass and very low mass stars of M35}
\begin{tabular}{cccrrrrrc}
\hline
I$_{\rm c}$ bin & (V--I)$_c$& (R--I)$_c$ & Contam. & 
                  N$_{i,*}^{\rm M35}$(I$_{\rm c}$)& Mass &
$\Delta$Mass & $\Delta$N/$\Delta$M & Dataset\\
  (1) &   (2)  &   (3) &   (4) &     (5)       &   (6)  &   (7)  & 
     (8)         &    (9)    \\
\hline
13.25 & 0.79 & --   &    0.40 & 37.9$\pm$04.4 &  1.356 & 0.138  & 275$\pm$031  &    kpno   \\    
13.75 & 0.85 & --   &    0.35 & 48.4$\pm$05.3 &  1.239 & 0.124  & 390$\pm$043  &    kpno   \\    
14.25 & 0.92 & --   &    0.30 & 52.2$\pm$10.0 &  1.120 & 0.110  & 474$\pm$091  &    kpno   \\    
14.75 & 1.02 & --   &    0.30 & 67.9$\pm$09.1 &  1.015 & 0.102  & 665$\pm$091  &    kpno   \\    
15.25 & 1.14 & --   &    0.30 & 64.0$\pm$07.0 &  0.915 & 0.097  & 659$\pm$072  &    kpno   \\    
15.75 & 1.29 & --   &    0.30 & 78.7$\pm$08.4 &  0.822 & 0.087  & 904$\pm$097  &    kpno   \\    
16.25 & 1.46 & --  &$\sim$0.30& 83.6$\pm$08.4 &  0.742 & 0.078  &1072$\pm$108  &    kpno   \\    
16.75 & 1.73 & --  &$\sim$0.30& 79.7$\pm$09.0 &  0.657 & 0.092  & 866$\pm$098  &    kpno   \\    
17.25 & 2.05 & --  &$\sim$0.30&100.3$\pm$11.2 &  0.569 & 0.077  &1303$\pm$146  &    kpno   \\    
17.75 & --   & 1.25&$\sim$0.02&115.5$\pm$21.2 &  0.497 & 0.075  &1540$\pm$283  &    cfht   \\    
18.25 & --   & 1.35&$\sim$0.02&138.0$\pm$17.5 &  0.417 & 0.085  &1624$\pm$206  &    cfht   \\    
18.75 & --   & 1.55&$\sim$0.02&187.5$\pm$21.5 &  0.330 & 0.085  &2206$\pm$253  &    cfht   \\    
19.25 & --   & 1.65&      0.07&168.9$\pm$17.3 &  0.252 & 0.069  &2448$\pm$251  &    cfht   \\    
19.75 & --   & 1.65&      0.35&166.3$\pm$23.9 &  0.191 & 0.052  &2819$\pm$460  &    cfht   \\    
20.25 & --   & 1.75&      0.60& 80.5$\pm$19.5 &  0.147 & 0.039  &2064$\pm$500  &    cfht   \\    
20.75 & --   & 1.85&      0.60& 58.5$\pm$09.0 &  0.112 & 0.031  &1887$\pm$290  &    cfht   \\    
21.25 & --   & 1.95&      0.60& 32.3$\pm$06.9 &  0.088 & 0.012  &2691$\pm$575  &    cfht   \\    
\hline
\end{tabular}
$\,$\\
%
%
\end{table*}

\end{document}